\documentclass[12pt]{iopjournal}
\usepackage{natbib}
\usepackage{booktabs}
\usepackage{longtable}
\usepackage{amsmath}
\usepackage{setspace}
\usepackage[utf8]{inputenc}
\begin{document}
\onehalfspacing

\articletype{Paper} 

\title{Teaching competencies in physics for engineering education: A qualitative analysis from teaching practice}

\author{Vanessa Cruz Molina$^1$, Daniel Sánchez Guzmán$^1$\orcid{0000-0001-9322-2734}, Teodoro Rivera Montalvo$^1$\orcid{0000-0002-6685-0687} and Ricardo García-Salcedo$^{1,*}$\orcid{0000-0003-0173-5466}}

\affil{$^1$Centro de Investigación en Ciencia Aplicada y Tecnología Avanzada, Unidad Legaria del Instituto Politécnico Nacional, Legaria 694, 11500, Ciudad de México, México.}


\affil{$^*$Author to whom any correspondence should be addressed.}

\email{rigarcias@ipn.mx}


\begin{abstract}
Physics teaching in engineering programmes poses discipline-specific demands that intertwine conceptual modelling, experimental inquiry, and computational analysis. This study examines nine teaching competences for physics instruction derived from international and regional frameworks and interpreted within engineering contexts. Nineteen university instructors from the Technological Institute of Toluca completed an open-ended questionnaire; responses were analysed using a grounded theory approach (open and axial coding) complemented by descriptive frequencies. Results indicate stronger development in technical mastery, methodological/digital integration, technology-mediated communication, and innovation (C1, C2, C6, C9), while information literacy for digital content creation/adaptation and digital ethics/safety (C7, C8) remain underdeveloped. A recurrent understanding–application gap was identified, revealing uneven transfer from conceptual awareness to enacted classroom practice. We conclude that advancing physics education for engineers requires institutionally supported, discipline-specific professional development that aligns modelling, laboratory work, and computation with ethical and reproducible digital practices; such alignment can move instructors from adoption/adaptation toward sustained appropriation and innovation in multimodal settings.
\end{abstract}

\section{Introduction}

Teaching physics in engineering programs presents discipline-specific challenges that go beyond generic pedagogical requirements. Physics is simultaneously a conceptual, mathematical, and experimental enterprise; effective instruction must therefore coordinate modelling, multiple representations (graphs, diagrams, vector fields, equations), measurement and uncertainty, and transfer to real technological contexts. Traditional transmissive approaches have proved insufficient to foster this coordination and to develop the analytical, experimental, and problem-solving abilities required of future engineers \citep{machin2017sostenibilidad, medina2011formacion, buils2023competencias}. These challenges have been amplified by the coexistence of face-to-face, hybrid, and online modalities, which demand from instructors not only disciplinary expertise but also robust pedagogical, digital, and socio-emotional competencies \citep{cabero2021digital, area2021ensenanza, salcines2018competencia}.

Aligned with international efforts to promote inclusive, equitable, and quality education, this study situates the development of physics-teaching competences within the framework of the 2030 Agenda for Sustainable Development, particularly Goal 4 (Quality Education) established by UNESCO. This goal emphasizes the need for higher education systems to foster scientific literacy, technological innovation, and sustainability-oriented practices. In this context, strengthening teachers’ disciplinary, digital, and pedagogical competencies in physics education is essential to ensure that future engineers acquire the critical and creative skills required to address contemporary global challenges \citep{unesco2017education2030}.

Over the last two decades, a competence-based perspective has guided higher-education reform, emphasizing the integration of knowing, doing, and being \citep{morin1999siete, perrenoud2004diez}. In physics education, this means that teachers must act as mediators of conceptual understanding, designers of experiment-rich learning environments, and reflective practitioners who can translate abstract formalism into empirically testable predictions. In engineering settings, effective physics teaching requires balancing mathematical modelling with laboratory inquiry and contextual application to real-world problems such as energy conversion, signal processing, fluid transport, or materials behaviour under load \citep{gomez2020perfil, torquemada2022evaluacion}. Thus, beyond generic teaching skills, physics instruction demands domain-specific competences that articulate theory, experiment, computation, and technological design.

The COVID-19 pandemic accelerated this competence-oriented shift. Emergency remote teaching exposed both the potential and the limitations of the academic community in integrating digital technologies into physics instruction (e.g., remote labs, simulations, data-acquisition apps, and computational notebooks) \citep{cabero2021digital, vargas2024competencia}. Many instructors demonstrated high disciplinary proficiency but limited digital and methodological competence, underscoring the need for targeted professional development in digital literacy, pedagogical innovation, and flexible instructional design aligned with physics tasks (modelling, error analysis, data fitting, and validation against physical constraints).

Recent studies in science and engineering education call for a systemic view of teaching competencies. Liu and Sun \citep{Liu2021} identify core competences for physics teachers—disciplinary, pedagogical, scientific, humanistic, informational, and lifelong learning—while Villarroel et al. \citep{villarroel2017competencias} propose macro-domains for engineering educators (professional awareness, theory–technology integration, social collaboration, methodological design). However, many frameworks remain generic and offer limited empirical evidence on how such competences are understood and enacted in physics classrooms.

The set of nine teaching competences considered in this study was informed by comparative analysis of international and regional frameworks—namely, the EU Key Competences for Lifelong Learning \citep{europaea2018key}, the European Commission’s DigCompEdu \citep{redecker2017european}, UNESCO’s ICT-CFT \citep{organizacion2019marco}, and Latin American adaptations \citep{zempoalteca2017formacion, fernandez2018competencias, guimaraes2022competencias}. A physics-specific interpretation guided the alignment with modelling, experimentation, and problem-solving in engineering contexts. (See Research Design and Methodology.)

This study addresses the empirical gap by characterizing the development of these nine teaching competences among university physics instructors in engineering programs, with explicit attention to their disciplinary enactment. In particular, C1 is interpreted as physics-specific proficiency: accurate use of canonical models and laws, facility with multiple representations and approximations, experimental reasoning (measurement, calibration, uncertainty), and the ability to apply these resources to ill-defined, real-world engineering problems. The remaining competences—methodological and digital integration, social interaction, assessment, adaptability and innovation, communication, resource management, professional reflection, and digital resilience—are likewise read through physics-education lenses (e.g., alignment of simulations and labs with learning goals; formative assessment of modelling assumptions; ethical and safe handling of experimental and digital data).

Methodologically, we draw on Grounded Theory \citep{strauss1998basics, corbin2014basics, charmaz2014constructing}, applying open and axial coding to open-ended responses from instructors. We identify levels of conceptual understanding and practical implementation for each competence, and we examine discipline-specific gaps between what instructors say they know about physics teaching and what they report doing in practice.

Our contribution is twofold. First, we offer an empirically grounded profile of physics-teaching competences situated in engineering education, making explicit how each competence manifests in tasks central to the discipline (modelling, experimentation, computation, and technological application). Second, we provide actionable evidence for professional development focused on didactic transfer—that is, turning disciplinary knowledge into practical, technology-enhanced, student-centred learning experiences in physics. This framing supports the ongoing transformation of physics education toward flexible, digitally supported, and context-aware instruction.


\section{Research Design and Methodology}

\subsection{Research approach}
The study adopts a qualitative–interpretive approach to understand how physics instructors in engineering programs conceptualize and enact teaching competencies in real educational settings \citep{denzin2011sage}. This paradigm is appropriate for exploring the meanings teachers attribute to their own practices. Descriptive quantitative procedures were also applied, particularly frequency counts of coded categories, to identify emerging trends and patterns within the data \citep{creswell2016qualitative}. This mixed (qualitative–quantitative) character allows for a richer interpretation of complex educational phenomena.
\subsection{Design}
For this research, a case study design was selected to provide an in-depth, contextually grounded analysis of the phenomenon \citep{yin2018case}. The case involves a cohort of university physics professors who teach in engineering programs at the Instituto Tecnológico de Toluca, a public technological university in Mexico. The case study design allows for examining the phenomenon within its institutional and disciplinary context, rather than seeking a statistical generalization \citep{merriam2015qualitative}.

The nine teaching competencies (C1–C9) included in this study were derived from an integrative synthesis of international and regional frameworks on teacher competence. From these references, nine competences were operationally defined: Technical mastery (C1); Methodological and digital integration (C2); Social interaction (C3); Assessment for learning (C4); Adaptability and innovation (C5); Communication (C6); Information management (C7); Digital ethics and safety (C8); and Professional reflection and resilience (C9). Their disciplinary interpretation aligned with physics education contexts (modelling, experimentation, data analysis, and problem-solving), ensuring that each competence reflected the specific demands of teaching physics in engineering programs. Finally, the proposed set was reviewed by three experts in higher education and physics teaching, whose feedback was used to validate the conceptual coherence and contextual relevance of the framework.

\subsection{Instrument}
An open-ended questionnaire was designed based on the nine teaching competences identified in the theoretical framework: technical, methodological, and digital, social, assessment, adaptability and innovation, communication, time and resource management, professional reflection, and digital resilience. Each competence was explored through two open questions:
\begin{enumerate}
    \item Q1: How do you understand this competence?
    \item Q2: How do you apply this competence in your teaching practice?
\end{enumerate}
This structure captured both the conceptual (understanding) and applied (implementation) dimensions of competence development \citep{cohen2002research}. Open-ended questions allowed participants to express their perspectives freely, producing rich qualitative data suitable for inductive analysis.

\subsection{Participants}
The analytical sample consisted of 19 university instructors who teach physics within engineering programs. Participants were selected intentionally according to the criterion of teaching at least one physics-related course. Table~\ref{tab:participants} summarizes the demographic profile of the sample.

\begin{table}[h]
\centering
\caption{Demographic and academic characteristics of the participants.}
\label{tab:participants}
\begin{tabular}{llll}
\toprule
\textbf{ID} & \textbf{Gender} & \textbf{Age range} & \textbf{Academic field} \\
\midrule
1 & Male & $>50$ & Electronics \\
2 & Male & 41--50 & Basic Sciences \\
3 & Female & 31--40 & Computer Systems \\
4 & Female & 41--50 & Mechatronics \\
5 & Male & 31--40 & Basic Sciences \\
6 & Female & 31--40 & Basic Sciences \\
7 & Male & 31--40 & Chemistry \\
8 & Male & 41--50 & Basic Sciences \\
9 & Female & 31--40 & Basic Sciences \\
10 & Female & $>50$ & Basic Sciences \\
11 & Male & $>50$ & Basic Sciences \\
12 & Female & 41--50 & Basic Sciences \\
13 & Male & 31--40 & Basic Sciences \\
14 & Male & $>50$ & Basic Sciences \\
15 & Male & 31--40 & Mechatronics \\
16 & Female & 41--50 & Basic Sciences \\
17 & Female & 31--40 & Basic Sciences \\
18 & Male & 31--40 & Basic Sciences \\
19 & Female & 31--40 & Basic Sciences \\
\bottomrule
\end{tabular}
\end{table}

The sample was balanced in gender (53\% male, 47\% female). Most instructors were between 31 and 50 years old and held degrees in Basic Sciences (68\%), followed by Mechatronics (11\%), Electronics (5\%), Chemistry (5\%), and Computer Systems (5\%). This composition reflects the interdisciplinary nature of physics teaching in engineering education. Although the results are not statistically generalizable, their transferability to comparable institutional contexts is plausible.

\subsection{Data collection}
Data were collected via a digital questionnaire implemented in Google Forms, open for two weeks during the 2024–2025 academic year. Participation was voluntary and anonymous. The online format enabled remote access and automatic export of data to spreadsheets for further analysis.

\subsection{Data analysis}
Data were analyzed through a thematic coding process inspired by Grounded Theory principles \citep{glaser2017discovery, strauss1998basics, charmaz2014constructing}. The analysis proceeded in several stages:
\begin{enumerate}
    \item Data preparation: All responses were compiled, standardized, and cleaned for coding.
    \item Open coding: Initial identification of emerging categories using a hybrid inductive–deductive strategy \citep{fereday2006demonstrating, braun2006using}. Two dimensions were distinguished: conceptual understanding and practical application.
    \item Axial coding: Categories were related to construct progressive levels of development along two axes: Understanding of the competence and Application in practice \citep{kitchener2006development, villarroel2017competencias}.
    \item Level grading: Five ordered levels were established for each axis, from superficial understanding to advanced application.
    \item Validation and cross-checking: Consistency between conceptual and applied dimensions was reviewed iteratively.
    \item Complementary quantitative analysis: Frequency counts by level were computed to reveal dominant patterns.
\end{enumerate}

Additionally, a gap analysis was conducted to assess the internal coherence between teachers’ conceptual understanding and practical implementation of each competence. Three analytical patterns were identified: (a) convergence (concept–practice alignment), (b) positive gap (conceptual awareness with low transfer), and (c) negative gap (practical implementation without explicit conceptual grounding). This comparison follows the constant-comparison logic of Grounded Theory and enhances interpretative rigor \citep{miles2014qualitative, bardin1991analisis}.

\subsection{Validity and reliability}
Several strategies were employed to ensure methodological rigor:
\begin{itemize}
    \item Triangulation: Cross-checking categories against specialized literature in physics education and teaching competences \citep{flick2015introducing}.
    \item Peer debriefing: Discussion of coding criteria among researchers to ensure internal consistency \citep{lincoln1985naturalistic}.
    \item Audit trail: Systematic documentation of analytical decisions, enhancing transparency and partial replicability \citep{miles2014qualitative}.
    \item Saturation criteria: Verification that category stability was reached before final interpretation.
\end{itemize}

This combined approach guarantees coherence and reliability in the findings, providing robust qualitative evidence on the teaching competencies of university physics instructors in engineering education.

\section{Results and Discussion}

\subsection{Overview of classification and competence levels}\label{sec:results-overview}
We classified open responses for the nine teaching competences across two dimensions: Conceptual Understanding and Application in Practice. Following the procedures outlined in the Methods, we used open and axial coding to derive five ordered levels in each dimension. For understanding: \textit{Null}, \textit{Low}, \textit{Medium}, \textit{High}, \textit{Very high}. For application: \textit{Access}, \textit{Adoption}, \textit{Adaptation}, \textit{Appropriation}, \textit{Innovation}. After coding all responses, absolute and relative frequencies were computed competence-by-competence. For interpretive clarity, \textit{Null}/\textit{Access}/\textit{Low} were, when relevant, grouped into a single initial operational tier in the application (''Access''). Table~\ref{TResumenRes} summarizes the distributions.

\begin{table}[h]
\centering
\caption{Percentage summary by competence (C1--C9) for \emph{Understanding} (Q1) and \emph{Application} (Q2), based on open-response coding ($n=19$).}

\label{TResumenRes}
\resizebox{\textwidth}{!}{
\begin{tabular}{lccccc|cccccc}
\toprule
 & \multicolumn{5}{c}{\textbf{Understanding \#(\%)}} & \multicolumn{5}{c}{\textbf{Application \#(\%)}} \\
\cmidrule(lr){2-6}\cmidrule(lr){7-11}
\textbf{Comp.} & Null & Low & Medium & High & Very high & Access & Adoption & Adaptation & Appropriation & Innovation \\
\midrule
C1 & 0 (0.0) & 5(26.3) & 8(42.1) & 4(21.1) & 2(10.5) & 0(0.0) & 5(26.3) & 9(47.4) & 5(26.3) & 0(0.0) \\
C2 & 1(5.3) & 0(0.0) & 7(36.8) & 9(47.4) & 2(10.5) & 1(5.3) & 8(42.1) & 2(10.5) & 8(42.1) & 0(0.0) \\
C3 & 4(21.1) & 2(10.5) & 7(36.8) & 5(26.3) & 1(5.3) & 7(36.8) & 3(15.8) & 4(21.1) & 4(21.1) & 1(5.3) \\
C4 & 3(15.8) & 1(5.3) & 10(52.6) & 5(26.3) & 0(0.0) & 6(31.6) & 3(15.8) & 4(21.1) & 6(31.6) & 0(0.0) \\
C5 & 0(0.0) & 2(10.5) & 9(47.4) & 8(42.1) & 0(0.0) & 2(10.5) & 6(31.6) & 4(21.1) & 7(36.8) & 0(0.0) \\
C6 & 1(5.3) & 0(0.0) & 6(31.6) & 12(63.2) & 0(0.0) & 3(15.8) & 10(52.6) & 4(21.1) & 2(10.5) & 0(0.0) \\
C7 & 9(47.4) & 0(0.0) & 5(26.3) & 4(21.1) & 1(5.3) & 10(52.6) & 2(10.5) & 4(21.1) & 3(15.8) & 0(0.0) \\
C8 & 12(63.2) & 0(0.0) & 4(21.1) & 3(15.8) & 0(0.0) & 12(63.2) & 1(5.3) & 3(15.8) & 3(15.8) & 0(0.0) \\
C9 & 5(26.3) & 0(0.0) & 3(15.8) & 10(52.6) & 1(5.3) & 6(31.6) & 1(5.3) & 1(5.3) & 8(42.1) & 3(15.8) \\
\bottomrule
\end{tabular}
}
\end{table}

\subsection{Competence-level patterns (C1–C9)}
The following competence–by–competence synthesis describes how instructors are distributed across ordered levels of \emph{Conceptual Understanding} and \emph{Application in Practice} (percentages over $n=19$; modes indicate central tendency, tails reflect heterogeneity). We interpret these distributions through physics–education lenses, i.e., in relation to core tasks of university physics for engineering: modelling and multiple representations, experimental design and uncertainty, data analysis and computation, and the contextualization of concepts to real technological problems. For each competence (C1–C9) we highlight (i) the modal level in understanding and application, (ii) the presence of convergences or gaps (understanding $\leftrightarrow$ application), and (iii) discipline–specific implications (e.g., transfer to laboratory practice, alignment of simulations with learning goals, or ethically sound handling of experimental/digital data). This reading frame prioritizes what these patterns \emph{mean for physics instruction}: where technical mastery supports modelling and experimentation (C1), where methodological and digital integration enables constructive alignment (C2, C6, C9), and where deficits in information literacy and digital ethics constrain rigorous, reproducible practice (C7, C8). We thus move beyond generic competence labels to their concrete enactment in physics teaching for engineering.

\paragraph{C1. Technical mastery.} Understanding concentrates at \textit{Medium} (42.1\%) with a tail to \textit{High/Very high} (31.6\%). Application peaks at \textit{Adaptation} (47.4\%), with \textit{Adoption} and \textit{Appropriation} at 26.3\% each. No cases of \textit{Access} or \textit{Innovation}. Overall: consolidated functional practice with room to progress towards systematic \textit{Appropriation}. This competence directly underpins students’ ability to translate fundamental laws of physics into quantitative models, simulations, and laboratory experiments relevant to engineering contexts.

\paragraph{C2. Methodological \& digital integration.} Understanding is \textit{High} (47.4\%) or \textit{Medium} (36.8\%); application is bimodal at \textit{Adoption}/\textit{Appropriation} (42.1\% each). Coherent discourse–practice, with a sizeable subset already planning technology-mediated activities purposefully. In physics teaching, this competence supports constructive alignment between theoretical lessons, computational modelling, and data-driven experimentation.

\paragraph{C3. Communicative competence.} While 68.4\% sit at \textit{Medium–High} understanding, application still clusters at \textit{Access} (36.8\%). This indicates a transfer gap from dialogic conceptions to dialogic classroom routines (\S\ref{sec:gap}). In the context of physics education, this limitation affects students’ capacity to reason collectively about models, argue from evidence, and articulate the physical meaning behind equations and data.

\paragraph{C4. Professional self-regulation.} Understanding \textit{Medium} (52.6\%) and \textit{High} (26.3\%); application splits between \textit{Access} and \textit{Appropriation} (31.6\% each). Heterogeneous trajectories: from declarative improvement to systematic reflective practice. In physics instruction, this competence is crucial for planning coherent laboratory sessions, managing uncertainty, and maintaining consistency between theoretical objectives and practical assessment.

\paragraph{C5. Information management.} Understanding \textit{Medium–High} (89.5\%); application led by \textit{Appropriation} (36.8\%). A minority operates at merely utilitarian search; most guide students towards curated sources, though advanced scholarly literacy is still uncommon. In physics, this competence is essential for guiding students in the critical evaluation of scientific data, proper citation of physical constants and datasets, and use of validated repositories for simulation and experimental analysis.

\paragraph{C6. Technology-mediated communication.} Understanding is \textit{High} (63.2\%); application mostly \textit{Adoption} (52.6\%), with fewer cases at \textit{Adaptation}/\textit{Appropriation}. Platforms are widely used, but bidirectional, structured interaction is less common. In the physics classroom, this competence enables the use of digital interfaces—such as virtual labs, sensors, and online simulations—to foster real-time feedback and collaborative exploration of physical phenomena.

\paragraph{C7. Digital content creation/adaptation.} Largest deficits: \textit{Null} understanding (47.4\%) and \textit{Access} application (52.6\%). A small emerging core reaches \textit{Appropriation} (15.8\%). This is a priority development area. In physics education, content creation is central to designing visualizations, interactive notebooks, and virtual experiments that make abstract models tangible to engineering students.

\paragraph{C8. Digital ethics \& safety.} Predominantly \textit{Null} understanding (63.2\%) and \textit{Access} application (63.2\%), with a minority at \textit{Adaptation/Appropriation}. Another critical development gap. In the physics laboratory, this competence is especially relevant for ensuring ethical data management, safe experimental handling, and respect for intellectual property when sharing digital materials or research outcomes.

\paragraph{C9. Innovation with ICT.} Understanding \textit{High} (52.6\%); application \textit{Appropriation} (42.1\%) with non-negligible \textit{Innovation} (15.8\%). Clear potential for leadership and scaling of innovative practice. Within physics education, this competence translates into the creative use of simulations, Arduino-based experiments, or open-source computational tools to promote inquiry-based learning and real-world engineering applications.

\subsection{Systematic profile synthesis}\label{sec:perfilesdocentes}
To integrate the two dimensions, each instructor was assigned a global profile (\textit{Incipient}, \textit{Intermediate}, \textit{Advanced}) using explicit correspondence rules based on predominant levels in understanding and application. Table~\ref{tab:regla_perfil} shows the mapping rule; Table~\ref{tab:perfil_calculo} reports individual assignments; Table~\ref{TPerfilesDocentes} provides brief qualitative descriptors per profile.

\begin{table}[ht]
\centering
\caption{Correspondence rule for global profile assignment.}
\label{tab:regla_perfil}
\begin{tabular}{lll}
\toprule
\textbf{Predominant understanding} & \textbf{Predominant application} & \textbf{Assigned profile} \\
\midrule
Null / Low          & Access / Adoption           & Incipient \\
Medium              & Adoption / Adaptation       & Intermediate \\
Medium / High       & Appropriation               & Intermediate \\
High / Very high    & Appropriation / Innovation  & Advanced \\
\bottomrule
\end{tabular}
\end{table}

\begin{longtable}{p{1cm}p{3.2cm}p{3.2cm}p{3cm}}
\caption{Predominant levels and assigned profile per instructor.}
\label{tab:perfil_calculo}\\
\toprule
\textbf{ID} & \textbf{Understanding} & \textbf{Application} & \textbf{Profile} \\
\midrule
\endfirsthead
\toprule
\textbf{ID} & \textbf{Understanding} & \textbf{Application} & \textbf{Profile} \\
\midrule
\endhead
1  & Medium       & Adaptation       & Intermediate \\
2  & Medium       & Appropriation    & Intermediate \\
3  & High         & Appropriation    & Advanced \\
4  & Low          & Access           & Incipient \\
5  & Medium       & Adoption         & Intermediate \\
6  & Medium       & Adoption         & Intermediate \\
7  & Low          & Access           & Incipient \\
8  & High         & Appropriation    & Advanced \\
9  & Medium       & Adaptation       & Intermediate \\
10 & Very high    & Appropriation    & Advanced \\
11 & Low          & Adoption         & Incipient \\
12 & High         & Appropriation    & Advanced \\
13 & Medium       & Adoption         & Intermediate \\
14 & Medium       & Adoption         & Intermediate \\
15 & Medium       & Access           & Incipient \\
16 & High         & Appropriation    & Advanced \\
17 & High         & Appropriation    & Advanced \\
18 & Medium       & Appropriation    & Intermediate \\
19 & Medium       & Adoption         & Intermediate \\
\bottomrule
\end{longtable}

{\tiny
\begin{longtable}{p{1cm} p{9cm} p{2.2cm}}
\caption{Qualitative characterization of profiles.}
\label{TPerfilesDocentes}\\
\toprule
\textbf{ID} & \textbf{Profile description} & \textbf{Profile} \\
\midrule
\endfirsthead
\caption*{\textit{(continued)}}\\
\toprule
\textbf{ID} & \textbf{Profile description} & \textbf{Profile} \\
\midrule
\endhead
1  & Medium understanding with consolidating application; functional resource use with punctual adjustments. & Intermediate \\
2  & Predominantly medium understanding with stable application; functional integration of ICT with moments of appropriation. & Intermediate \\
3  & High to very high understanding; appropriation with occasional innovation. & Advanced \\
4  & Many gaps and instrumental use; limited didactic integration. & Incipient \\
5  & Mixed (medium--high) understanding; mainly adoption with context-sensitive adjustments. & Intermediate \\
6  & Low--medium understanding; basic/fragmentary application. & Incipient \\
7  & Medium--high understanding; adoption--adaptation with punctual material production. & Intermediate \\
8  & High/very high understanding; systematic appropriation with emerging innovation. & Advanced \\
9  & Solid conceptual grasp; still routine application; scope for reflective improvement. & Intermediate \\
10 & Integral understanding; consolidated application with active methods and innovation. & Advanced \\
11 & Low--medium understanding; active application in exercises/projects; needs stronger didactic integration. & Intermediate \\
12 & High/very high understanding; reflective application with planning and ethical awareness. & Advanced \\
13 & Partial to medium understanding; punctual innovation and material creation; on the way to appropriation. & Intermediate \\
14 & Medium--high understanding; functional adoption--adaptation with selective ICT use. & Intermediate \\
15 & Medium understanding; basic application with examples; moving toward methodological adjustments. & Intermediate \\
16 & High self-awareness and understanding; consistent appropriation with interactive resources. & Advanced \\
17 & Medium--high understanding; appropriation in assessment/materials; punctual ICT leadership. & Intermediate \\
18 & High understanding in key foci; reflective application via PBL/projects; some gaps remain. & Intermediate \\
19 & Medium understanding; mostly instrumental application with some adjustments. & Intermediate \\
\bottomrule
\end{longtable}
}

\subsection{Association with sociodemographic variables}
Given the small sample ($n=19$), analyses are exploratory. Spearman’s $\rho$ for \textit{Age--Profile} was weak and non-significant: $\rho=0.131$, $p=0.593$. Chi-square tests indicate independence between \textit{Gender} and \textit{Profile}, and between \textit{Academic field} and \textit{Profile}, with Cramér’s $V$ suggesting non-significant moderate effects, especially for academic field ($V=0.46$). See Table~\ref{tab:chi_cramer}.

\begin{table}[ht]
\centering
\caption{$\chi^2$ and Cramér’s $V$ for nominal variables ($n=19$).}
\label{tab:chi_cramer}
\begin{tabular}{lcccc}
\toprule
\textbf{Variable pair} & $\chi^2$ & df & $p$ & Cramér’s $V$ \\
\midrule
Gender -- Profile        & 3.46  & 2 & 0.178 & 0.35 \\
Field of study -- Profile& 12.06 & 8 & 0.148 & 0.46 \\
\bottomrule
\end{tabular}
\end{table}

\subsection{Gap analysis: understanding vs application}\label{sec:gap}
We contrasted modal levels in both dimensions for each competence using ordinal codings (\textit{Understanding}: 0–4; \textit{Application}: 0–4) and $\Delta=\text{Application}_{\text{mode}}-\text{Understanding}_{\text{mode}}$. Criteria: $\Delta=0$ convergence; $|\Delta|=1$ mild gap; $|\Delta|\ge2$ marked gap. Table~\ref{TabBrecha} summarizes results.

\begin{table}[ht]
\centering
\caption{Gap analysis by competence (modal levels, difference, interpretation).}
\label{TabBrecha}
\resizebox{\textwidth}{!}{
\begin{tabular}{lccccp{8.5cm}}
\toprule
\textbf{Comp.} & \textbf{Understanding (mode)} & \textbf{Application (mode)} & \textbf{$\Delta$} & \textbf{Type} & \textbf{Synthesis} \\
\midrule
C1 & Medium (42.1\%) & Adaptation (47.4\%) & 0 & Convergent & Mid-level conceptual base with contextual adjustments; no Innovation. \\
C2 & High (47.4\%) & Adoption/Appropriation (42.1/42.1\%) & -- & Mixed/polarized & Coexistence of basic and advanced practice under high understanding. \\
C3 & Medium (36.8\%) & Access (36.8\%) & $-2$ & Marked positive gap & Knowing $>$ doing: dialogic concepts not yet internalized in routines. \\
C4 & Medium (52.6\%) & Access/Appropriation (31.6/31.6\%) & -- & Mixed/polarized & Elementary practices co-exist with consolidated appropriation. \\
C5 & Medium (47.4\%) & Appropriation (36.8\%) & $+1$ & Mild negative gap & Doing $>$ saying: tacit know-how surpasses explicit conceptualization. \\
C6 & High (63.2\%) & Adoption (52.6\%) & $-2$ & Marked positive gap & Strong conceptual grasp with limited transfer to advanced interaction. \\
C7 & Null (47.4\%) & Access (52.6\%) & 0 & Low-level convergence & Structural deficit in content design/adaptation. \\
C8 & Null (63.2\%) & Access (63.2\%) & 0 & Critical convergence & Deficit in ethics/safety: priority for intervention. \\
C9 & High (52.6\%) & Appropriation (42.1\%) & 0 & High convergence & Coherent discourse–practice; some Innovation (15.8\%). \\
\bottomrule
\end{tabular}}
\end{table}

\subsection{Cross-level synthesis}
To reveal global tendencies, levels were normalized into three operational bands: \textit{Low} (Null/Low/Access), \textit{Medium} (Medium/Adoption/Adaptation), \textit{High} (High/Very high/Appropriation/Innovation). Aggregated distributions indicate Medium–High understanding (41.0\% / 37.3\%) vs Medium–Low application (41.6\% / 34.8\%). Thus, conceptual mastery generally exceeds operational enactment. Competences with stronger integral development are C6 and C9; C7 and C8 remain the lowest in both dimensions and warrant targeted development.

\subsection{General discussion}

Three cross-cutting insights emerge from the analysis, each shedding light not only on the patterns observed but also on the structural factors that help explain them.

\begin{enumerate}
\item Solid conceptual bases with uneven transfer.
Most instructors articulate contemporary pedagogical ideas (C2, C6, C9), yet classroom routines often remain at \textit{Adoption}/\textit{Adaptation}. Advancing to \textit{Appropriation}/\textit{Innovation} requires constructive alignment (intended outcomes–activities–assessment) and sustained formative feedback. However, the data suggest that the limited transfer from conceptual awareness to enacted practice may stem from contextual constraints: lack of time for instructional redesign, scarce institutional incentives to experiment with active methodologies, and the absence of a shared curricular framework that mandates reflective planning. In physics education, these barriers are compounded by the heavy cognitive and logistical load of coordinating theory, laboratory work, and computational modelling, which often discourages systematic innovation despite teachers’ willingness to update their approaches.

\item Critical digital literacies lag.
Digital content design/adaptation (C7) and ethics/safety (C8) exhibit low understanding and application, consistent with known gaps in higher education. These weaknesses are not only individual but systemic. Many physics instructors rely on inherited teaching materials or institutionally provided platforms, limiting opportunities to design their own interactive or ethical digital resources. Furthermore, institutional training programs tend to emphasize technical skills (platform use) over epistemic and ethical dimensions of digital competence. This imbalance restricts teachers’ ability to guide students in responsible data handling and reproducible experimentation. Professional development initiatives should therefore integrate the design and ethical evaluation of digital physics materials as a core component of teacher formation.

\item Profiles concentrate at \textit{Intermediate}.
The profile synthesis reveals a teaching community in transition: instructors exhibit adequate conceptual discourse, but practices are still in the process of consolidation. This pattern points to a structural equilibrium in which professional growth occurs primarily through self-directed reflection rather than through formal institutional processes. Advancing from intermediate to advanced profiles will likely depend on establishing structured mentoring, collaborative design studios, and recognition mechanisms for pedagogical innovation in physics. The creation of disciplinary communities of practice—where instructors exchange experimental setups, simulation codes, and inquiry-based tasks—could foster the sustained diffusion of innovation across programs and institutions.
\end{enumerate}

Thus, the competence gaps observed are not solely cognitive but also organizational and systemic. Addressing them requires institutional policies that allocate time, incentives, and shared frameworks for pedagogical innovation, ensuring that physics teaching in engineering moves from isolated good practices toward a coherent, research-informed model of professional development.

\subsection{Implications and limitations}

From a disciplinary standpoint, three implications are salient for physics teaching in engineering programs. 
(i) Focus faculty development on didactic transfer within core physics practices. Professional learning should explicitly align conceptual aims with enacted tasks typical of the discipline (modelling with multiple representations, laboratory inquiry with error analysis and calibration, and computation for simulation and data fitting), so that instructors move from \textit{Adoption/Adaptation} to \textit{Appropriation/Innovation} in C1--C2 and C6--C9. 
(ii) Embed scholarly information literacy and digital ethics across physics coursework. Curricular and mentoring structures should address C7--C8 through activities that make responsible data handling, reproducibility (e.g., notebooks and versioning), and informed use of simulations/AI tools a visible part of laboratory reports, projects, and assessments. 
(iii) Institutionalize collaborative design and feedback cycles. Departments should establish communities of practice that share laboratory templates, simulation repositories, rubrics for model validation, and peer-observation protocols, coupled with recognition mechanisms and workload protection, to stabilize advanced practice over time.

The study is limited by its small, single-case sample ($n=19$) and by the self-reported nature of the responses, which may introduce desirability bias. These risks were mitigated through explicit coding criteria, peer debriefing, and an audit trail, yet the absence of classroom observations or artefact analysis constrains ecological validity. Future research should extend this work through multi-site and longitudinal designs that link teachers’ competence profiles with student learning outcomes in modelling, experimentation, and computation.

Taken together, these implications frame the concluding claim of this paper: advancing physics education in engineering requires not only teachers’ conceptual awareness but also institutionally supported, discipline-specific mechanisms that translate competencies C1--C9 into coherent sequences of modelling, experimentation, and computation. The \emph{Conclusions} synthesize how our evidence substantiates this claim and delineates actionable priorities for departments and programs.

\section{Conclusions}

This study examined the development of teaching competencies among university physics instructors in engineering programs, emphasizing their disciplinary role in fostering conceptual understanding, modelling skills, and experimental reasoning through pedagogically grounded, technology-supported instruction. Using a mixed–methods approach informed by grounded theory and descriptive frequency analysis, the research explored both the conceptual and applied dimensions of teachers’ professional performance in physics education.

Findings reveal that participants are at an intermediate stage of competence development, with solid disciplinary and pedagogical awareness but uneven translation into enacted classroom practice. Higher levels of understanding were found in competencies linked to technical mastery (C1), methodological and digital integration (C2), communication (C6), and innovation with ICT (C9). Conversely, competencies related to information literacy and digital ethics (C7, C8) remain underdeveloped. Overall, while most instructors recognize the pedagogical value of technology, their use of it in physics teaching often remains instrumental—supporting exposition rather than enabling genuine inquiry, modelling, or experimentation.

The comprehension–application gap observed across competences highlights the challenge of \emph{didactic transfer}—transforming disciplinary knowledge of physics into meaningful learning experiences that bridge theoretical models with real-world engineering applications. In particular, competence C1 (\emph{Technical mastery}) transcends conceptual familiarity with physical laws: it demands the ability to model mechanical, electrical, or thermodynamic systems, interpret measurement uncertainties, and connect laboratory or simulated data to design solutions. Similarly, C2 (\emph{Methodological and digital integration}) is essential for creating inquiry-based activities where computational tools, data-acquisition systems, or modelling software mediate between theoretical and empirical reasoning. Thus, the development of physics-teaching competencies must be understood as the capacity to integrate disciplinary reasoning with pedagogical and technological mediation.

From an institutional perspective, three key implications arise:

\begin{enumerate}
    \item \textbf{Competency-based professional development anchored in physics practice.}  
    Faculty training should explicitly link pedagogical strategies to the epistemic practices of physics—modelling, experimentation, and data interpretation—so that conceptual mastery translates into authentic, inquiry-oriented learning experiences.

    \item \textbf{Reinforcing digital and ethical literacy in physics instruction.}  
    Beyond technical proficiency, teachers require critical awareness of data reliability, digital safety, and ethical management of experimental and computational results. These elements are integral to scientific integrity and to the responsible formation of engineers in the digital era.

    \item \textbf{Building communities of reflective physics educators.}  
    Collaborative environments in which instructors share laboratory designs, simulation codes, and assessment rubrics can foster sustained pedagogical innovation and the collective construction of knowledge in physics teaching.
\end{enumerate}

Beyond the diagnostic scope of this study, the empirical patterns identified suggest specific pathways for developing competencies among physics teachers in engineering. For competencies C1-C2, structured workshops on the constructive alignment of modeling, simulation, and experimentation could strengthen didactic transfer. For competencies C3-C6, microteaching and peer feedback practices can foster the internalization of dialogic and assessment-oriented routines. For competencies C7-C8, institutional programs focused on digital content creation, data integrity, and the responsible use of computational tools are required. Finally, competency C9 (professional reflection and resilience) could be enhanced through mentoring programs and communities of practice that promote collaborative lesson design and reflective documentation. While this study did not measure the impact of such interventions, these recommendations outline a viable agenda for their future implementation and evaluation.

Although limited by sample size and the self-reported nature of responses, the study privileges interpretive depth and methodological rigor through systematic coding, triangulation, and peer validation. Future research should pursue longitudinal and multi-site studies to track the evolution of disciplinary competencies after targeted interventions and develop hybrid instruments that combine quantitative indicators with open qualitative analysis, thereby correlating conceptual understanding with applied teaching practice.

In conclusion, this research provides an empirically grounded diagnosis of physics-teaching competencies within engineering education. The results depict a teaching community in transition—conceptually mature in disciplinary understanding yet still evolving in its ability to operationalize that knowledge through inquiry-based, technology-enhanced, and ethically responsible practices. Strengthening the alignment between physics content, pedagogical innovation, and engineering application will be essential to advance toward a coherent, research-informed model of physics education for the twenty-first century.

\section*{Acknowledgements}
We thank the Secretaría de Investigación y Posgrado of the Instituto Politécnico Nacional for funding this work (projects SIP20230505, SIP20240638, and SIP20250043) and the FORDECYT-PRONACES-CONACYT project CF-MG-2558591. We also acknowledge the use of artificial intelligence tools to support English translation and stylistic editing of this manuscript, under the full supervision and verification of the authors.

\section*{Data availability statement}

The datasets generated and analyzed during the current study consist of qualitative and quantitative responses from university instructors. For ethical and privacy reasons, these data are not publicly available. However, anonymized excerpts or aggregated data can be made available from the corresponding author upon reasonable request.

\bibliographystyle{unsrtnat}
\bibliography{CTesisV}

\begin{thebibliography}{36}
\providecommand{\natexlab}[1]{#1}
\providecommand{\url}[1]{\texttt{#1}}
\expandafter\ifx\csname urlstyle\endcsname\relax
  \providecommand{\doi}[1]{doi: #1}\else
  \providecommand{\doi}{doi: \begingroup \urlstyle{rm}\Url}\fi

\bibitem[Machín~Armas et~al.(2017)Machín~Armas, Céspedes~Montano, Riverón~Mena, and Fernández~Santiesteban]{machin2017sostenibilidad}
Francisco~O. Machín~Armas, Santiago~G. Céspedes~Montano, Aleida~N. Riverón~Mena, and Eduardo Fernández~Santiesteban.
\newblock Sostenibilidad, ingeniería y enseñanza de las ciencias básicas. marco teórico conceptual.
\newblock \emph{Revista Iberoamericana de Educación}, 73:\penalty0 179--202, 2017.
\newblock \doi{10.35362/rie730298}.

\bibitem[Medina~Rivilla et~al.(2011)Medina~Rivilla, Dom{\'\i}nguez~Garrido, and Ribeiro~Gon{\c{c}}alves]{medina2011formacion}
Antonio Medina~Rivilla, M{\textordfeminine}~Dom{\'\i}nguez~Garrido, and Fernando Ribeiro~Gon{\c{c}}alves.
\newblock Formaci{\'o}n del profesorado universitario en las competencias docentes.
\newblock \emph{Revista historia de la educaci{\'o}n latinoamericana}, 13\penalty0 (17):\penalty0 119--138, 2011.

\bibitem[Buils et~al.(2023)Buils, Arroyo-Ainsa, S{\'a}nchez-Tarazaga, Esteve-Mon, et~al.]{buils2023competencias}
Sara Buils, Patricia Arroyo-Ainsa, Luc{\'\i}a S{\'a}nchez-Tarazaga, Francesc~M Esteve-Mon, et~al.
\newblock Competencias docentes para el desarrollo profesional en la universidad actual.
\newblock \emph{Journal of Supranational Policies of Education}, .\penalty0 (17):\penalty0 76--102, 2023.

\bibitem[Cabero-Almenara et~al.(2021)Cabero-Almenara, Guill{\'e}n-G{\'a}mez, Ruiz-Palmero, and Palacios-Rodr{\'\i}guez]{cabero2021digital}
Julio Cabero-Almenara, Francisco~D Guill{\'e}n-G{\'a}mez, Julio Ruiz-Palmero, and Antonio Palacios-Rodr{\'\i}guez.
\newblock Digital competence of higher education professor according to digcompedu. statistical research methods with anova between fields of knowledge in different age ranges.
\newblock \emph{Education and Information Technologies}, 26\penalty0 (4):\penalty0 4691--4708, 2021.

\bibitem[Area-Moreira(2021)]{area2021ensenanza}
Manuel Area-Moreira.
\newblock La enseñanza remota de emergencia durante la covid-19. los desafíos postpandemia en la educación superior.
\newblock \emph{Propuesta Educativa}, 56:\penalty0 57--70, 2021.
\newblock URL \url{https://propuestaeducativa.flacso.org.ar/wp-content/uploads/2022/04/REVISTA-56-Dossier-AREA-MOREIRA.pdf}.

\bibitem[Salcines-Talledo(2018)]{salcines2018competencia}
Irina Salcines-Talledo.
\newblock La competencia digital docente: análisis de su inclusión en los planes de estudio de las titulaciones de educación.
\newblock \emph{Revista Complutense de Educación}, 29\penalty0 (2):\penalty0 335--354, 2018.

\bibitem[{UNESCO}(2017)]{unesco2017education2030}
{UNESCO}.
\newblock \emph{Education for Sustainable Development Goals: Learning Objectives}.
\newblock UNESCO Publishing, Paris, 2017.
\newblock URL \url{https://unesdoc.unesco.org/ark:/48223/pf0000247444}.

\bibitem[Morin(1999)]{morin1999siete}
Edgar Morin.
\newblock Los siete saberes necesarios para la educaci{\'o}n del futuro, 1999.

\bibitem[Perrenoud(2004)]{perrenoud2004diez}
Philippe Perrenoud.
\newblock \emph{Diez nuevas competencias para enseñar}.
\newblock Graó, Barcelona, 2004.

\bibitem[G{\'o}mez~Jim{\'e}nez et~al.(2020)G{\'o}mez~Jim{\'e}nez, Ram{\'\i}rez~D{\'\i}az, and Arriaga~Santos]{gomez2020perfil}
Irene G{\'o}mez~Jim{\'e}nez, Mario~Humberto Ram{\'\i}rez~D{\'\i}az, and Carlos~Adri{\'a}n Arriaga~Santos.
\newblock El perfil del docente de f{\'\i}sica como factor en el desarrollo de las competencias del estudiante en el bachillerato.
\newblock \emph{RIDE. Revista iberoamericana para la investigaci{\'o}n y el desarrollo educativo}, 11\penalty0 (21), 2020.

\bibitem[Torquemada~Gonz{\'a}lez(2022)]{torquemada2022evaluacion}
Alma~Delia Torquemada~Gonz{\'a}lez.
\newblock Evaluaci{\'o}n, desarrollo, innovaci{\'o}n y futuro de la docencia universitaria. de la red iberoamericana de investigadores en evaluaci{\'o}n de la docencia.
\newblock \emph{Perfiles educativos}, 44\penalty0 (175):\penalty0 200--204, 2022.

\bibitem[Vargas-Franco et~al.(2024)Vargas-Franco, L{\'o}pez-Gil, and Torres-Casierra]{vargas2024competencia}
Alfonso Vargas-Franco, Karen L{\'o}pez-Gil, and Liana~Mercedes Torres-Casierra.
\newblock La competencia digital docente en tiempos de pandemia: actitudes, pr{\'a}cticas y procesos formativos.
\newblock \emph{Tecn{\'e}, Episteme y Didaxis: TED}, Jan/Jun 2024\penalty0 (55):\penalty0 226--245, 2024.

\bibitem[Liu and Sun(2021)]{Liu2021}
X.~Liu and L.~Sun.
\newblock A framework of physics teacher competence for quality education.
\newblock \emph{Journal of Baltic Science Education}, 20\penalty0 (3):\penalty0 432--445, 2021.

\bibitem[Villarroel and Bruna(2017)]{villarroel2017competencias}
Ver{\'o}nica~A Villarroel and Daniela~V Bruna.
\newblock Competencias pedag{\'o}gicas que caracterizan a un docente universitario de excelencia: un estudio de caso que incorpora la perspectiva de docentes y estudiantes.
\newblock \emph{Formaci{\'o}n universitaria}, 10\penalty0 (4):\penalty0 75--96, 2017.

\bibitem[Europaea(2018)]{europaea2018key}
Schola Europaea.
\newblock Key competences for lifelong learning in the european schools.
\newblock \emph{Office of the secretary-general of the European schools, pedagogical development unit: Brussels, Belgium}, 72, 2018.

\bibitem[Redecker and Punie(2017)]{redecker2017european}
Christine Redecker and Yves Punie.
\newblock European framework for the digital competence of educators: Digcomp edu.
\newblock \emph{Luxembourg, Luxembourg: Publications Office of the European Union. https://doi. org/10.2760/159770}, 2017.

\bibitem[Organizaci{\'o}n de las Naciones Unidas para~la Educaci{\'o}n(2019)]{organizacion2019marco}
la~Ciencia y la~Cultura Organizaci{\'o}n de las Naciones Unidas para~la Educaci{\'o}n.
\newblock Marco de competencias de los docentes en materia de tic.
\newblock \emph{UNESCO}, 2019.
\newblock URL \url{https://www.unesco.org/en/digital-competencies-skills/ict-cft}.

\bibitem[Zempoalteca~Dur{\'a}n et~al.(2017)Zempoalteca~Dur{\'a}n, Barrag{\'a}n~L{\'o}pez, Gonz{\'a}lez~Mart{\'\i}nez, and Guzm{\'a}n~Flores]{zempoalteca2017formacion}
Beatriz Zempoalteca~Dur{\'a}n, Jorge~Francisco Barrag{\'a}n~L{\'o}pez, Juan Gonz{\'a}lez~Mart{\'\i}nez, and Teresa Guzm{\'a}n~Flores.
\newblock Formaci{\'o}n en tic y competencia digital en la docencia en instituciones p{\'u}blicas de educaci{\'o}n superior.
\newblock \emph{Apertura (Guadalajara, Jal.)}, 9\penalty0 (1):\penalty0 80--96, 2017.

\bibitem[Fern{\'a}ndez-M{\'a}rquez et~al.(2018)Fern{\'a}ndez-M{\'a}rquez, Leiva-Olivencia, and L{\'o}pez-Meneses]{fernandez2018competencias}
Esther Fern{\'a}ndez-M{\'a}rquez, Juan~Jos{\'e} Leiva-Olivencia, and Eloy L{\'o}pez-Meneses.
\newblock Competencias digitales en docentes de educaci{\'o}n superior.
\newblock \emph{Revista digital de investigaci{\'o}n en docencia universitaria}, 12\penalty0 (1):\penalty0 213--231, 2018.

\bibitem[Guimaraes et~al.(2022)Guimaraes, Aroca, Mart{\'\i}nez, Re{\'a}tegui, and V{\'a}squez]{guimaraes2022competencias}
Jos{\'e} Lisbinio~Cruz Guimaraes, Brigitte Elizabeth~Llantoy Aroca, Manuel Jes{\'u}s~Guevara Mart{\'\i}nez, Andy Walter~Rivera Re{\'a}tegui, and Ang{\'e}lica Mar{\'\i}a~Minchola V{\'a}squez.
\newblock Competencias digitales de docentes en la educaci{\'o}n superior universitaria: retos y perspectivas en el {\'a}mbito de la educaci{\'o}n virtual.
\newblock \emph{Ciencia Latina Revista Cient{\'\i}fica Multidisciplinar}, 6\penalty0 (1):\penalty0 1536--1567, 2022.

\bibitem[Strauss and Corbin(1998)]{strauss1998basics}
Anselm Strauss and Juliet Corbin.
\newblock \emph{Basics of qualitative research techniques}.
\newblock Thousand oaks, CA: Sage publications, 1998.

\bibitem[Corbin and Strauss(2014)]{corbin2014basics}
Juliet Corbin and Anselm Strauss.
\newblock \emph{Basics of qualitative research: Techniques and procedures for developing grounded theory}.
\newblock Sage publications, 2014.

\bibitem[Charmaz(2014)]{charmaz2014constructing}
Kathy Charmaz.
\newblock \emph{Constructing grounded theory}.
\newblock SAGE publications Ltd, 2014.

\bibitem[Denzin and Lincoln(2011)]{denzin2011sage}
Norman~K Denzin and Yvonna~S Lincoln.
\newblock \emph{The Sage handbook of qualitative research}.
\newblock sage, 2011.

\bibitem[Creswell and Poth(2016)]{creswell2016qualitative}
John~W Creswell and Cheryl~N Poth.
\newblock \emph{Qualitative inquiry and research design: Choosing among five approaches}.
\newblock Sage publications, 2016.

\bibitem[Yin(2018)]{yin2018case}
Robert~K Yin.
\newblock \emph{Case study research and applications}, volume~6.
\newblock Sage Thousand Oaks, CA, 2018.

\bibitem[Merriam(2015)]{merriam2015qualitative}
Sharan~B Merriam.
\newblock Qualitative research: Designing, implementing, and publishing a study.
\newblock In \emph{Handbook of research on scholarly publishing and research methods}, pages 125--140. IGI Global Scientific Publishing, 2015.

\bibitem[Cohen et~al.(2002)Cohen, Manion, and Morrison]{cohen2002research}
Louis Cohen, Lawrence Manion, and Keith Morrison.
\newblock \emph{Research methods in education}.
\newblock routledge, 2002.

\bibitem[Glaser and Strauss(2017)]{glaser2017discovery}
Barney Glaser and Anselm Strauss.
\newblock \emph{Discovery of grounded theory: Strategies for qualitative research}.
\newblock Routledge, 2017.

\bibitem[Fereday and Muir-Cochrane(2006)]{fereday2006demonstrating}
Jennifer Fereday and Eimear Muir-Cochrane.
\newblock Demonstrating rigor using thematic analysis: A hybrid approach of inductive and deductive coding and theme development.
\newblock \emph{International journal of qualitative methods}, 5\penalty0 (1):\penalty0 80--92, 2006.

\bibitem[Braun and Clarke(2006)]{braun2006using}
Virginia Braun and Victoria Clarke.
\newblock Using thematic analysis in psychology.
\newblock \emph{Qualitative research in psychology}, 3\penalty0 (2):\penalty0 77--101, 2006.

\bibitem[Kitchener et~al.(2006)Kitchener, King, and DeLuca]{kitchener2006development}
Karen~Strohm Kitchener, Patricia~M King, and Sonia DeLuca.
\newblock Development of reflective judgment in adulthood.
\newblock \emph{Handbook of adult development and learning}, pages 73--98, 2006.

\bibitem[Miles et~al.(2014)Miles, Huberman, Salda{\~n}a, et~al.]{miles2014qualitative}
Matthew~B Miles, A~Michael Huberman, Johnny Salda{\~n}a, et~al.
\newblock \emph{Qualitative data analysis: A methods sourcebook. 3rd}.
\newblock Thousand Oaks, CA: Sage, 2014.

\bibitem[Bardin(1991)]{bardin1991analisis}
Laurence Bardin.
\newblock \emph{An{\'a}lisis de contenido}, volume~89.
\newblock Ediciones Akal, 1991.

\bibitem[Flick(2015)]{flick2015introducing}
Uwe Flick.
\newblock \emph{Introducing research methodology: A beginner's guide to doing a research project}.
\newblock Sage, 2015.

\bibitem[Lincoln and Guba(1985)]{lincoln1985naturalistic}
Yvonna~S Lincoln and Egon~G Guba.
\newblock \emph{Naturalistic inquiry}.
\newblock sage, 1985.

\end{thebibliography}

\end{document}